\documentclass[%
preprint,
superscriptaddress,
 amsmath,amssymb,
 aps,
]{revtex4-2}

\usepackage{graphicx}
\usepackage{dcolumn}
\usepackage{bm}
\usepackage{hyperref}
\hypersetup{
    colorlinks=true,
    urlcolor= blue,
    citecolor=blue,
linkcolor= blue}

\newcommand{\CdAs}{$\mathrm{Cd_{3}As_{2}}$}
\newcommand{\Vg}{$V_{\mathrm{g}}$}
\newcommand{\rxx}{$\rho_{xx}$}
\newcommand{\rxy}{$\rho_{xy}$}
\newcommand{\sxx}{$\sigma_{xx}$}
\newcommand{\sxy}{$\sigma_{xy}$}
\newcommand{\VgO}{$V_{\mathrm{G}}^{\mathrm{CNP}}$}

\usepackage{amssymb}   
\usepackage{bm}        
\usepackage{dsfont}   
\usepackage{mathrsfs}    
\usepackage{amsmath}   

\DeclareMathOperator{\BFk}{\mathbf{k}}

\def\be{\begin{equation}} \def\ee{\end{equation}}
\def\bea{\begin{eqnarray}} \def\eea{\end{eqnarray}}

\def\BFk{{\mathbf{k}}}

\usepackage{hyperref}
\usepackage{chemformula}

\usepackage{amsthm}    

\usepackage{graphicx}

\usepackage{float}

\begin{document}

\preprint{APS/123-QED}

\title {Integer quantum Hall effect and enhanced g-factor in quantum confined \texorpdfstring {Cd$_3$As$_2$}{Cd3As2} films}
\author{Run Xiao}
\affiliation{Department of Physics, The Pennsylvania State University, University Park, PA 16802, USA}
\author{Junyi Zhang}
\affiliation{Institute for Quantum Matter and Department of Physics and Astronomy, Johns Hopkins University, Baltimore, Maryland 21218, USA}
\author{Juan Chamorro}
\affiliation{Department of Chemistry, Johns Hopkins University, Baltimore, Maryland 21218, USA}
\author{Jinwoong Kim}
\affiliation{Department of Physics and Astronomy, Rutgers University, Piscataway, New Jersey 08854 USA}
\author{Tyrel M. McQueen}
\affiliation{Department of Chemistry, Johns Hopkins University, Baltimore, Maryland 21218, USA}
\author{David Vanderbilt}
\affiliation{Department of Physics and Astronomy, Rutgers University, Piscataway, New Jersey 08854 USA}
\author{Morteza Kayyalha}
\affiliation{Department of Electrical Engineering, The Pennsylvania State University, University Park PA 16802}
\author{Yi Li}
\affiliation{Institute for Quantum Matter and Department of Physics and Astronomy, Johns Hopkins University, Baltimore, Maryland 21218, USA}
\author{Nitin Samarth}
\thanks{Corresponding author: nsamarth@psu.edu}
\affiliation{Department of Physics, The Pennsylvania State University, University Park, PA 16802, USA}

\date{\today}

\begin{abstract}
We investigate the integer quantum Hall effect in \CdAs~thin films under conditions of strong to moderate quantum confinement (thicknesses of 10 nm, 12 nm, 15 nm). In all the films, we observe the integer quantum Hall effect in the spin-polarized lowest Landau level (filling factor $\nu = 1$) and at spin-degenerate higher index Landau levels with even filling factors ($\nu = 2,4,6$). With increasing quantum confinement, we also observe a lifting of the Landau level spin degeneracy at $\nu  = 3$, manifest as the emergence of an anomaly in the longitudinal and Hall resistivity. Tight-binding calculations show that the enhanced g-factor likely arises from a combination of quantum confinement and corrections from nearby subbands. We also comment on the magnetic field induced transition from an insulator to a quantum Hall liquid when the chemical potential is near the charge neutrality point.
\end{abstract}

\maketitle

Over the past decade, the interplay between strong spin-orbit coupling (SOC) and crystalline symmetry has created a rich playground for the study of topological phases of quantum matter in materials. Examples include topological insulators \cite{hasan2010colloquium,qi2011topological} and topological Dirac and Weyl semimetals \cite{armitage2018weyl}. Measurements of quantum transport in such materials yield important insights into the unusual electronic states that host these topological phases \cite{konig2007quantum,chang2013experimental,liang2015ultrahigh,xiong2015evidence,Huang_PhysRevX.5.031023,liang2018experimental,ong2021experimental,bernevig2022progress}. In this broad context, the archetypal Dirac semimetal \CdAs~\cite{wang2013three,liu2014discovery}, with space group $I4_1/acd$ and two Dirac points on the $\Gamma$-$Z$ path protected by inversion, time-reversal symmetry, and $C_{4}$ rotational symmetry~\cite{Ali2014b,Baidya2020}, provides a model platform for understanding the underlying topological bulk and surface states (Dirac cones and Fermi arcs, respectively) via measurements of the integer quantum Hall effect (IQHE)  \cite{schumann2018observation,uchida2017quantum,Goyal_APLMater,Galletti_PhysRevB.99.201401,Zhang_Weyl_Orbit_Nature,Zhang_Weyl_Orbit_NC2017,uchida2017quantum,nishihaya2018gate,nishihaya2019quantized,nishihaya2021intrinsic}. In the fully three-dimensional (3D) regime (sample thickness $\sim 80 - 150 ~\mathrm{nm}$), measurements of the IQHE in \CdAs~nanoplates have been interpreted in terms of Weyl orbits that involve surface Fermi arcs on opposite surfaces \cite{Zhang_Weyl_Orbit_NC2017,Zhang_Weyl_Orbit_Nature,li20203d}. The situation becomes more complex in thinner \CdAs~films (sample thickness $\sim 10 - 30 ~\mathrm{nm}$). In this regime, quantum confinement is expected to affect the 3D bulk states via a topological phase transition that accompanies the opening of a finite gap between the conduction subbands and valence subbands. At the same time, the gapless surface states on the top and bottom surfaces of the {\CdAs} thin film can hybridize and open a hybridization gap (as in ultrathin topological insulator films \cite{Liu2010}).  Thus, understanding the IQHE in this quantum confined regime of \CdAs~films requires insights into the full interplay between the 2D quantum confined bulk states \cite{wang2013three}, the hybridized topological surface states (closed surface Fermi pockets) \cite{Kargarian8648,villanova2017engineering} and surface Fermi arcs \cite{arribi2020topological,potter2014quantum}. Despite concerted efforts to distinguish these various possibilities in experiments \cite{uchida2017quantum,schumann2018observation,Goyal_APLMater,Galletti_PhysRevB.99.201401,nishihaya2018gate,nishihaya2019quantized,nishihaya2021intrinsic}, a complete understanding of the IQHE in \CdAs~films in the strongly quantum confined regime (thinner than about 20 nm) remains to be satisfactorily resolved \cite{Galletti_PhysRevB.99.201401,Chen_PhysRevResearch.3.033227}. A full first-principles calculation of the Landau levels (LLs) in this regime is also challenging, especially for the most commonly studied (112)-oriented \CdAs~films.   

Here, we report on a concerted experimental and theoretical effort to understand the IQHE in \CdAs~thin films in the limit of strong to moderate quantum confinement (film thickness $10 ~\mathrm{nm} \leq t \leq 15 ~\mathrm{nm}$). We map out the LL spectrum via measurements of the Shubnikov de Haas (SdH) quantum oscillations, and of the fully developed IQHE, as a function of chemical potential and magnetic field ($B \leq 9$~T) at low temperature ($T = 50$ mK). In \CdAs~films with thickness of 15 nm, apart from the $\nu = 1$ quantum Hall plateau, we only observe plateaus at even filling factors $\nu = 2,4,6$. This is similar to earlier reports of IQHE in \CdAs~films of comparable thickness \cite{schumann2018observation,uchida2017quantum}. In the thinnest sample that we studied (10 nm thick), we observe the emergence of the $\nu = 3$ quantum Hall state, suggesting that the spin degeneracy has been removed at a relatively modest magnetic field ($B = 9$~T). As an aside, we note that prior quantum transport measurements in very high magnetic fields ($B \geq 15$~T) have shown spin-resolved quantum Hall states with both even and odd filling factors in thicker, quasi-3D \CdAs~films \cite{uchida2017quantum}; this is likely the result of the large g-factor in bulk \CdAs~crystals, with varying estimated values of $g \sim 16$ \cite{Narayanan_PhysRevLett.114.117201} and $g \sim 37$ \cite{Jeon_NatMat_2014}. Our observations are consistent with a new understanding of LLs that emerges from four-band tight-binding model calculations carried out for [112]-oriented quantum confined \CdAs~films. Our calculations show that quantum confinement results in Landau levels (LLs) derived from both bulk subbands and surface states. This produces a non-monotonic dependence of the g-factor on thickness so that an enhanced Zeeman splitting is possible when the film thickness is decreased from 15 nm to 10 nm. This prediction is consistent with the emergence of a $\nu = 3$ quantum Hall state in the two thinner samples measured in our experiments. Finally, we briefly address quantum transport in \CdAs~thin films when the chemical potential is close to the charge neutrality point (CNP). Here, reduced Coulomb screening leads to a localized insulating phase at zero magnetic field. The application of a magnetic field then leads to the observation of a striking direct transition from a trivial insulator to the $\nu = 1$ or $\nu  = 2$ quantum Hall liquid states without any precursor SdH oscillations. We propose that studies of the IQHE in this regime provide a new Dirac material platform to revisit the insulator to quantum Hall liquid transition, originally studied in semiconductor-based 2D electron gases \cite{Jiang_PhysRevLett.71.1439,Song_PhysRevLett.78.2200,Smorchkova_PhysRevB.58.R4238} but rarely reported in Dirac materials such as graphene \cite{Huang_C6RA07859A}.   

We grew [112]-oriented \CdAs~thin films by molecular beam epitaxy (MBE) in a Veeco EPI 930 chamber using epi-ready miscut GaAs (111)B substrates ($1^{\circ}$ toward $(2\bar{1}\bar{1})$). The epi-ready GaAs substrates were first annealed inside the MBE chamber to flash off the native oxide at a thermocouple temperature of 820 $^{\circ}$C ($T_{\rm{actual}} \sim 580^{\circ}$C) while monitoring the surface using 12 keV reflection high energy electron diffraction (RHEED). Then, we deposited a thin ($\sim 2 $nm) GaAs layer at the same substrate temperature to smoothen the surface. Subsequently, the substrates were cooled down to thermocouple temperature of 520$^{\circ}$C) under As$_4$ flux for the growth of a 100 nm GaSb buffer layer with Sb/Ga beam equivalent pressure (BEP) ratio of 7. The substrates were then cooled down further to 400 $^{\circ}$C under Sb$_4$ flux, and further cooled down to a thermocouple temperature of 180 $^{\circ}$C after closing the Sb shutter. Once the sample temperature was stable at 180 $^{\circ}$C ($T_{\rm{actual}} \sim 110^{\circ}$C), we evaporated a high-purity {Cd$_3$As$_2$} compound source from a standard Knudsen effusion cell to deposit a \CdAs~thin film. The desired BEP flux and the growth rate of \CdAs~were controlled by the effusion cell temperature. A BEP flux of around $8 \mu$Pa yielded a growth rate of about 0.33 nm/min. During the growth, the pressure of the MBE chamber was maintained at $2 \times 10^{-8}$ Pa.
We used high resolution cross-sectional transmission electron microscopy (HRTEM) to accurately determine the crystalline structure and the thickness (with an uncertainty of $\pm 1$ nm) of all films for which we report measurements in this paper \cite{Xiao_supp}. Finally, we also used X-ray diffraction (XRD) and determined that only peaks from \CdAs $(112)$ planes are observed. We also note that we have previously used angle resolved photoemission spectroscopy (ARPES) measurements to show the presence of the expected Dirac semimetal band structure in \CdAs~films grown under similar conditions in the same MBE system \cite{Yanez_PhysRevApplied.16.054031}.  All these characterization measurements confirm that our films have the correct distorted anti-fluorite phase of \CdAs~for forming a 3D Dirac semimetal. 

After the sample growth, we used photolithography and Argon plasma etching to pattern the films into two different device geometries: 50 $\mu \mathrm{m} \times 25 ~\mu \mathrm{m}$ Hall bars and 25 $\mu \mathrm{m} \times 25 ~\mu \mathrm{m}$ Van der Pauw configurations. We then evaporated 10 nm Ti/30 nm Au as contact electrodes. Finally, we fabricated a top gate comprised of a 30 nm {Al$_2$O$_3$} dielectric layer and 10 nm Ti/30 nm Au contacts, deposited using atomic layer deposition and electron beam evaporation, respectively. Transport measurements were carried out in a He3/He4 dilution refrigerator (Blue Fors, base temperature $T < 50$ mK, $B \leq 9$ T) using lock-in techniques with an excitation current of 1 nA. At base temperature and zero gate voltage, the samples were n-type due to the naturally formed As vacancies, with carrier density of about $1 \times 10^{12}$ cm$^{-2}$ and Drude mobility $\sim 1 \times 10^{4}$ cm$^{2}$/V.s. The application of a voltage to the top gate (\Vg) allows the carrier density in the samples to be continuously tuned from n-type to p-type.

\begin{figure*}
\includegraphics[width=0.9\textwidth]{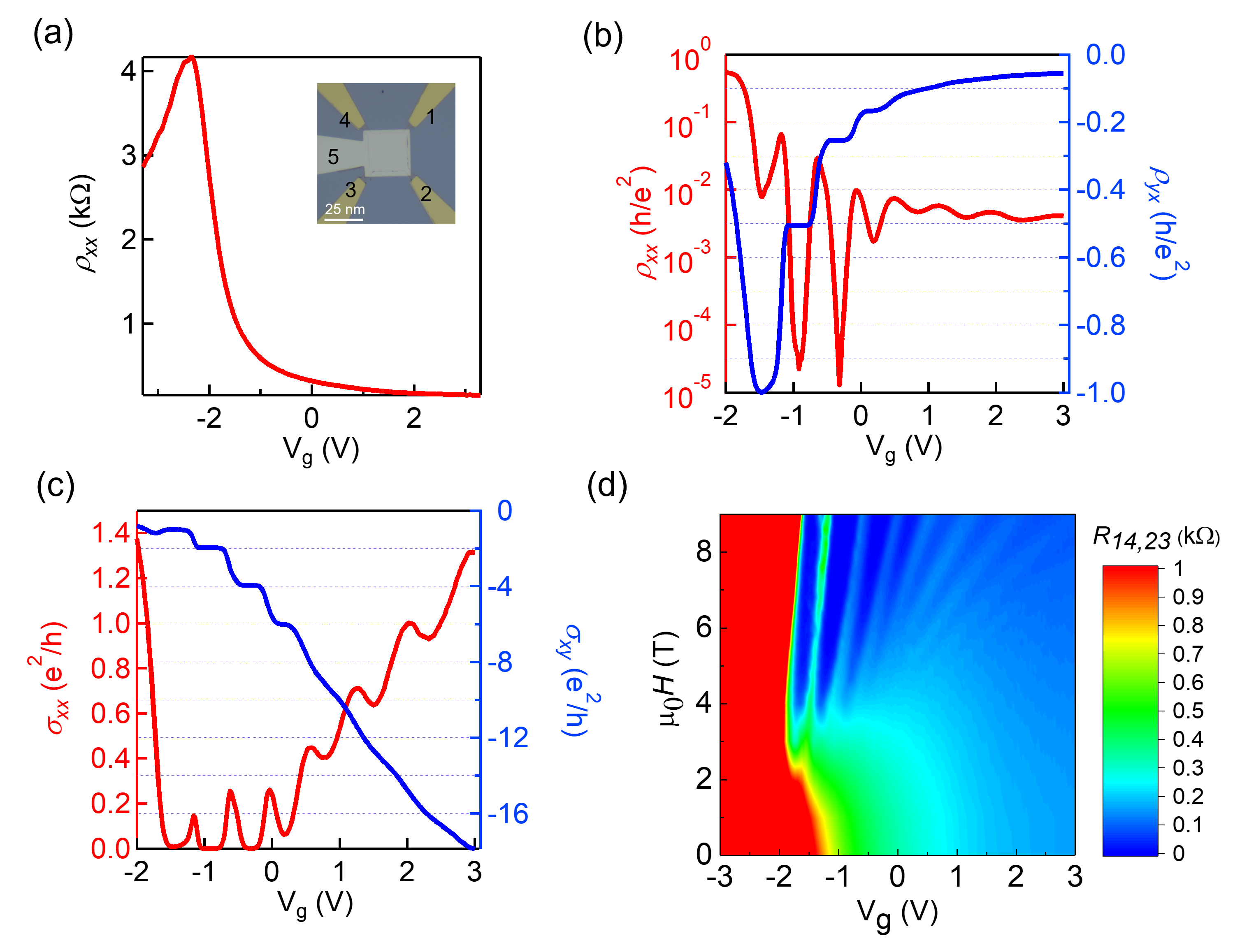}
\caption{\label{Figure 1}IQHE in a moderately quantum confined 15 nm thick {Cd$_3$As$_2$} film at $T = 50$ mK. (a) Gate-voltage dependence of \rxx~at zero magnetic field. The inset shows a schematic of the Van der Pauw geometry.  (b) Gate voltage dependence of \rxx~(red, left axis) and \rxy~(blue, right axis) at $B = 9$~T, showing quantized Hall plateaus at filling factor $\nu=1,2,4,6$ accompanied by deep minima in \rxx. The finite minimum at $\nu = 1$ (\rxx $\sim 200~\Omega$) is attributed to the incomplete development of the first LL at $B = 9$~T. (c) Gate voltage dependence of \sxx~(red, left axis) and \sxy~(blue, right axis) at $B = 9$~T. These values are obtained from the data in (b). (d) Variation of $R_{14,23}$ with $B$ and \Vg, where $R_{14,23} = V_{23}/I_{14}$. The plot is generated by sweeping \Vg~at fixed values of $B$. Magnetic field is incremented in steps of 0.2 T for $B \geq 4$~T and in steps of 1 T for $B < 4$~T. The data are then smoothly interpolated. A raw plot is provided in the Supplementary material \cite{Xiao_supp}.
}
\end{figure*}

We first discuss the observation of the IQHE in a 15 nm thick \CdAs~film patterned into a van der Pauw geometry device with an electrostatic top gate. The IQHE can be measured either by varying the chemical potential at fixed magnetic field or by varying the magnetic field as fixed gate voltage. In the latter case, the value of the chemical potential changes with magnetic field to accommodate the increasing degeneracy of Landau levels. We chose the former approach for convenience. All values of the longitudinal resistivity (\rxx) and Hall resistivity (\rxy) that we report have been properly averaged for the van der Pauw configuration. The magnetic field $B$ is always applied perpendicular to the plane of the sample. Figure \ref{Figure 1} (a) shows the variation of \rxx~as a function of \Vg~at zero magnetic field. At \Vg$= 0$V, the sample is n-type (see details in Supplementary Materials \cite{Xiao_supp}). Applying a negative gate voltage moves the chemical potential toward the valence band and \rxx~rises to a peak value of 4.2 k$\Omega$ at the charge neutral point (CNP) at \Vg$=$\VgO $= -2$V. Applying a positive gate voltage shifts the chemical potential higher up into the conduction band and reduces \rxx~from its value at zero gate voltage. Figure \ref{Figure 1} (b) shows the variation of \rxx~and \rxy~with \Vg~at a fixed field magnitude $B = 9$T. We see clear evidence of the IQHE wherein Hall plateaus (\rxy $= \frac{h}{\nu e^2}$) are accompanied by \rxx$\sim 0$. By using the standard conversion of \rxx~into a Hall conductivity \sxy, we find that the filling factors associated with these plateaus are $\nu = 1, 2, 4, 6$ (Fig. \ref{Figure 1} (c)). For the quantum Hall plateaus with $\nu = 1, 2, 4$ in Fig.\ref{Figure 1} (b), $\rho_{xx}$ has values  203.6  $\Omega$,  0.92 $\Omega$,  0.33 $\Omega$, and $\rho_{yx}$ has values 0.997, 0.4967, and 0.2472 h/e$^{2}$, respectively. It is interesting to note that the $\nu = 1$ quantum Hall state is not as robust as $\nu = 2$: the plateau is not as wide in the former as it is in the latter and the minimum in \rxx~is also not as deep. This indicates that the gap at $\nu = 1$ is smaller than that associated with $\nu = 2$. This resembles the behavior seen in single layer and bilayer graphene.    

Since the IQHE is not well-developed at large filling factors ($\nu > 6$),  we determine the filling factors using the derivative of $\rho_{xx}$. First, Hall plateaus and quantum oscillations corresponding to higher filling factors are easier to determine in the derivative of $\rho_{xx}$ (see Fig S1 in the Supplementary Materials \cite{Xiao_supp}). This yields even filling factors $\nu = 8, 10$. Next, we construct a Landau level (LL) fan diagram by plotting \rxx~as a function of $B$ and \Vg. Figure \ref{Figure 1} (d) shows such a plot for a given orientation of magnetic field. The plot is generated by sweeping \Vg~at fixed magnetic field and the quantity plotted is $R_{14,23}$ wherein current is passed between terminals 1,2 and voltage is measured across terminals 3,4 (see inset in Fig. \ref{Figure 1} (a)). The linear-in-field features in the LL fan can be assigned to filling factors $\nu= 1, 2, 4, 6, 8, 10, 12$ (see Figure S2 of the Supplementary Materials \cite{Xiao_supp}). We thus conclude that the IQHE in 15 nm \CdAs~thin films only occurs at $\nu = 1$ and at higher even filling factors; this is consistent with previous studies of \CdAs~thin films of similar thickness ($t \sim 10 - 20$~nm)  \cite{uchida2017quantum,schumann2018observation,Goyal_APLMater}. However, as we show below, our measurements of the IQHE in the 10 nm \CdAs~film reveal a new feature not seen in previous reports.

\begin{figure*}
\includegraphics[width=0.9\textwidth]{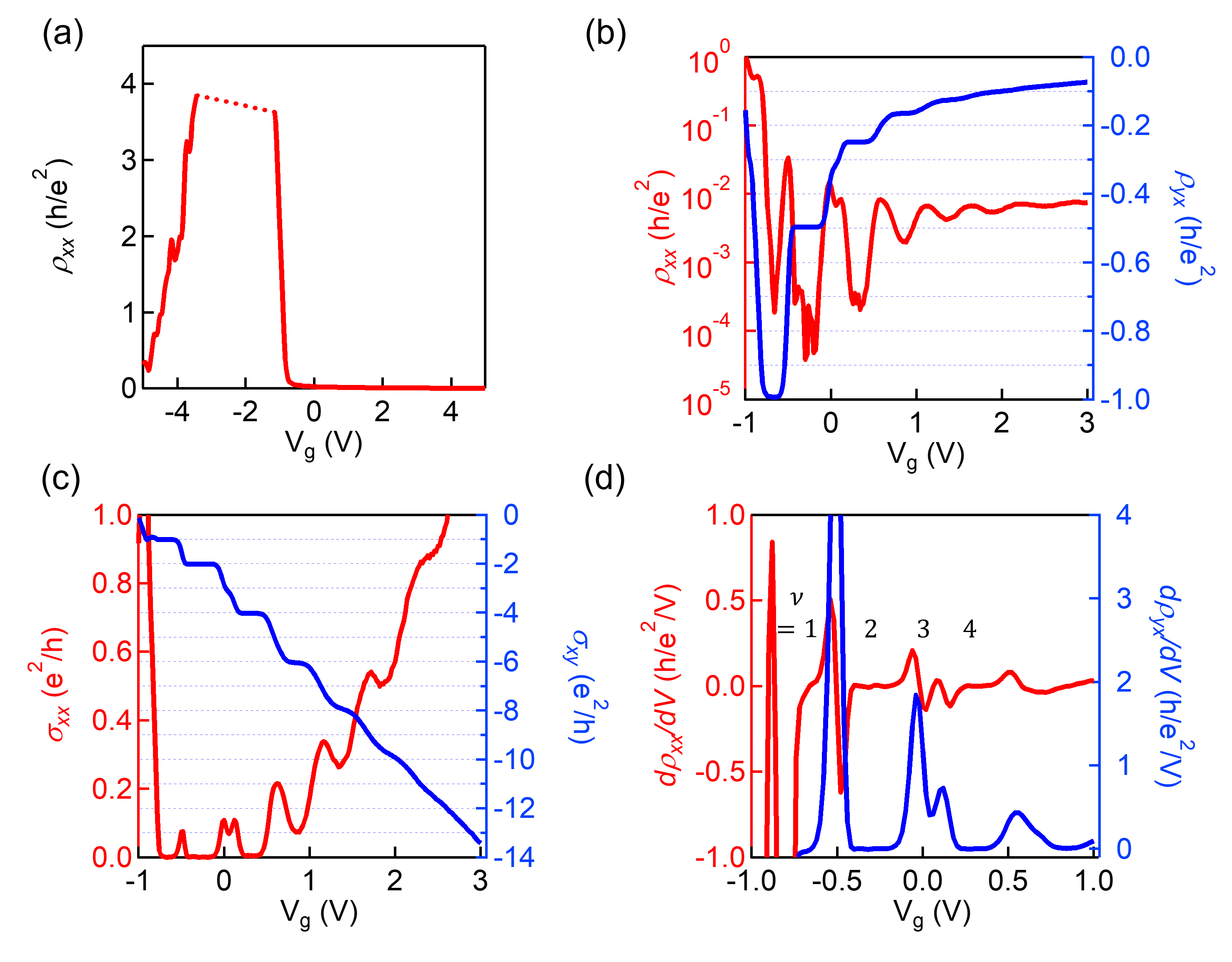}
\caption{\label{Figure 2}IQHE in a strongly quantum confined 10 nm thick {Cd$_3$As$_2$} thin film at $T = 50$ mK. (a) Variation of \rxx~with \Vg~at $B = 0$. When the chemical potential is close to the CNP, we cannot reliably measure \rxx~using lockin techniques (represented by the dashed line). (b) \Vg-dependence of \rxx~(red, left axis) and \rxy~(blue, right axis) at $B = 9$ T. For the $\nu = 1, 2, 4$ quantum Hall plateaus, $\rho_{xx}$ has values of 4.8 $\Omega$, 0.98 $\Omega$,  5.9 $\Omega$, while $\rho_{yx}$ has values 0.9937, 0.4973, and 0.2483 h/e$^{2}$, respectively. (c) \Vg-dependence of \sxx~(red, left axis) and \sxy~(blue, right axis) at $B = 9$ T. These values are obtained from the data in (b). (d) First derivative plot of the data in (b), showing the quantum Hall states with filling factor $\nu= 1,2,3,4$.
}
\end{figure*}

\begin{figure*}
\includegraphics[width=0.9\textwidth]{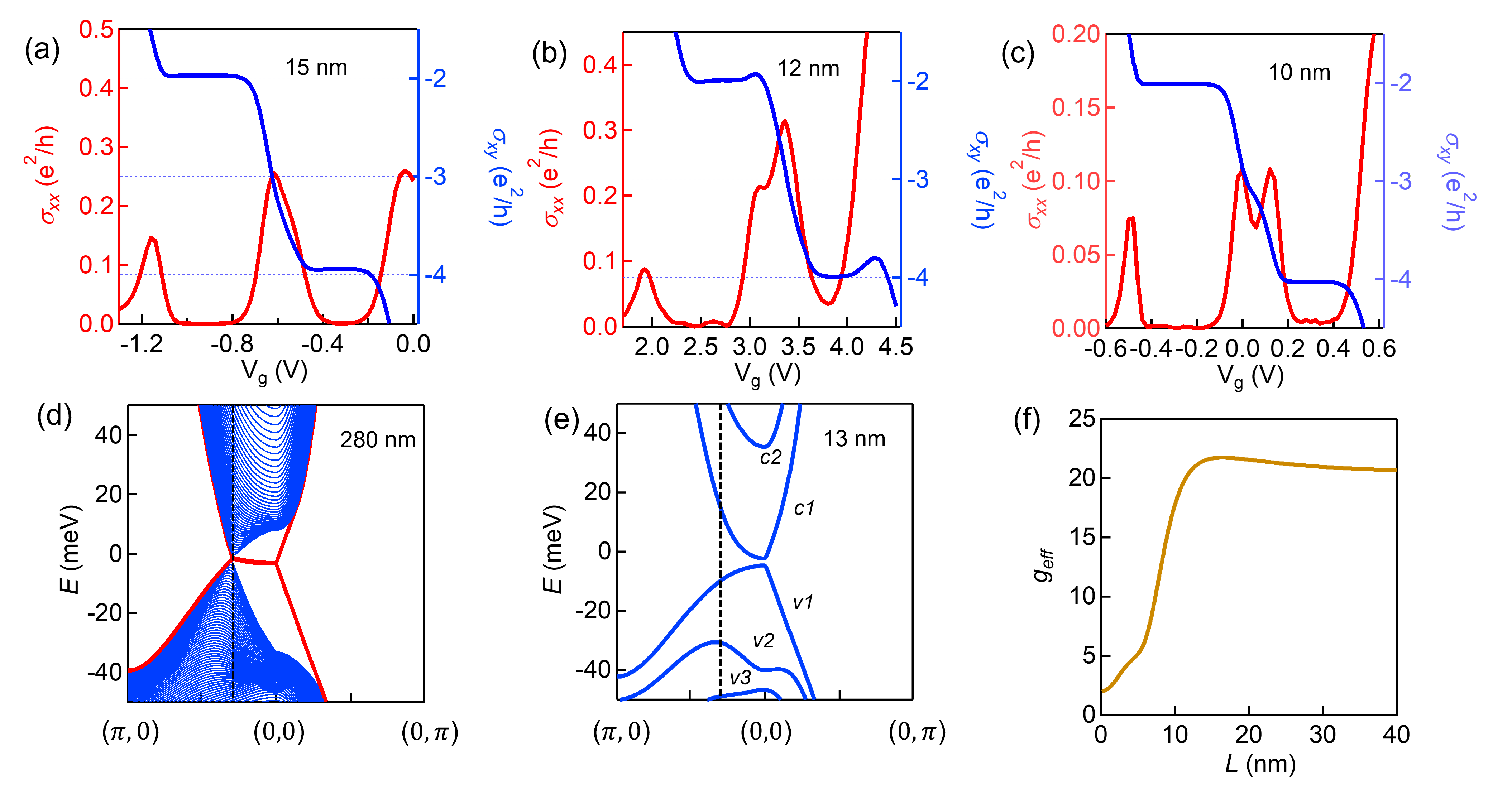}
\caption{\label{Figure 3} (a)-(c) Experimentally measured evolution of the $\nu=3$ feature in \CdAs~films with different thickness: (a) 15 nm,  (b) 12 nm, (c) 10 nm. Each panel shows the \Vg-dependence of \sxx~(red, left axis) and \sxy~(blue, right axis) at $B = 9$ T. (d)-(e) Band structure in the surface Brillouin zone of $[112]$-oriented films, based on a tight-binding model, of two different thicknesses: (d) 280 nm (e) 13 nm. The vertical dashed line indicates the projection of the bulk Dirac point $\mathbf{k}_D^+$ in the SBZ at $(0.3\pi,0)$. Blue and red indicate bulk and surface states, respectively, in panel (d). The calculation shows a quantum confinement induced band gap of 2.3 meV in the thinner case. (f) Calculated g-factors as a function of {Cd$_3$As$_2$} film thickness.
}
\end{figure*}

Figure \ref{Figure 2} shows the transport properties of a 10 nm thick \CdAs~thin film at 50 mK. At $B = 0$ T, \rxx~dramatically increases when the chemical potential is tuned to the CNP, with the resistance of the sample going beyond the range of the lock-in amplifier, showing a completely insulating behavior. At $B = 9$~T, when the chemical potential is tuned away from the CNP, the sample exhibits the IQHE, similar to the 15 nm thick film, with quantum Hall plateaus corresponding to $\nu = 1,2,4,6,8$ (Figs. \ref{Figure 2} (b),(c)). In addition, we observe the signature of the $\nu = 3$ quantum Hall state by fine tuning \Vg. This is seen more clearly in the first derivative plots of \rxx~and\rxy~with respect to \Vg~when \Vg $\approx 0.06$~V (Fig. \ref{Figure 2} (d)). In Fig. \ref{Figure 3} (a)- (c), we compare the evolution of the signature of the $\nu = 3$ quantum Hall state with different thicknesses. Note that there is no sign of the $\nu=3$ state in the 15 nm thick film; this only becomes more obvious as the quantum confinement becomes stronger in the two thinner films.

We now seek a theoretical understanding of the quantum confinement effects on the band structure and how the modifications in band structure affect the LLs and the g-factor in ultrathin {\CdAs} films. 
The quantum confinement gives rise to quasi-2D subbands derived from 3D electronic states.
As the bulk Dirac points located at $\mathbf{k}_{D}^{\pm} = \pm(0,0,k_D)$ are projected to the (112) surface Brillouin zone (SBZ) to two separate points, $\bar{\mathbf{k}}_{D}^+$ and $\bar{\mathbf{k}}_{D}^-$, surface (or surface-like) states are supported in (112)-oriented films.  Figure \ref{Figure 3} (d) shows the projected bulk bands (blue) and the surface states (red) for a $280$nm thick film, calculated using a four-band tight-binding model~\cite{Zhang2022} (see also Sec. V in the Supplementary Materials~\cite{Xiao_supp}). The existence of surface states has also been confirmed by ARPES experiments in bulk crystals~\cite{yi2014evidence,roth2018reinvestigating} and more recently in 30 nm thin films \cite{Yanez_PhysRevApplied.16.054031}. 
In thick films, the surface Dirac node at $\bar{\Gamma}$ (the projection of $\Gamma$ point in SBZ) is topologically protected, while the dispersion of the surface states depends on many factors such as the confinement potential, nature of the substrate, and the termination layer. As chemical potential varies, a ring-shaped closed Fermi arc around $\bar{\Gamma}$ or a pair of open Fermi arcs  bridging $\bar{\BFk}_D^+$ and $\bar{\BFk}_D^-$ can be supported. The topology of the surface Fermi arcs changes at a surface Lifshitz transition~\cite{Zhang2022}.

In thin films, quantum confinement strongly modifies the band structure. It splits bulk bands into quasi-2D subbands and can further hybridize topological gapless surface states discussed above to form gapped surface-like bands. 
As shown in Fig. \ref{Figure 3} (e), for the tight-binding model band structure of a $(112)$-oriented {\CdAs} film with thickness $t\sim 13$nm, as an example, 
the energy separation between the subbands near $\bar{\mathbf{k}}_{D}^{\pm}$ is of the order of $30$ meV. This is consistent with the confinement  splitting $\Delta E ={\hbar \pi v}/{t}$ where $\hbar v \sim 0.1 \text{eV}\cdot \mathrm{nm}$ near $\mathbf{k}_{D}^{\pm}$ \cite{Baidya2020}. 
In addition to the confinement splitting, Fig. \ref{Figure 3} (e) shows a smaller gap,  $\Delta_{hyb}\sim 2.3$meV, between the lowest conduction subband $c_1$, and the highest valence subband $v_1$, near $\bar{\Gamma}$. 
Comparing with Fig. \ref{Figure 3} (d), we can see that the dispersions of $c_1$ and $v_1$ near $\bar{\Gamma}$ are modified slightly from those of the surface states in the thicker film. 
Furthermore, $\Delta_{hyb}$ increases for thinner films, so $\Delta_{hyb}$ arises from the hybridization of surface states. 

Since the IQHE is observed in ultrathin films of {\CdAs} with low $n$-doped carrier concentrations, we focus on the contribution from the lowest conduction subband $c_1$. 
As $c_1$ hybridizes with the highest valence subband $v_1$ near $\bar{\Gamma}$, 
these two bands together can be described by an anisotropic 2D massive Dirac Hamiltonian \begin{equation}\label{eq:MDBand_MassiveDiracHamiltonian}
\begin{split}
    H_{MD} = \Delta_{hyb} \sigma_z +  v_\xi \Pi_{\xi} \sigma_x + v_\eta \Pi_{\eta} \sigma_y,
\end{split}
\end{equation} 
under a magnetic field along $[112]$, where 
$v_i$ and  $\Pi_i = p_i + eA_i$ are respectively the Dirac velocity and kinetic momentum along the $i$-th direction in the $(112)$ plane, with $i=\xi$ and $\eta$~\footnote{The in-plane $\xi$-axis is along the projection of $c$-axis and the in-plane $\eta$-axis is perpendicular to $\xi$-axis. In SBZ, the $(\pi, 0)$-direction ($(0, \pi)$ resp.) is along $\xi$-axis ($\eta$-axis resp.). }.

We solve the Landau level energies of the above anisotropic massive Dirac subbands as~\cite{Xiao_supp}
\begin{equation}\label{eq:MDBand_FormulaLLSpectrumFitting}
\begin{split}
    E_{LL}^{n_L} =& \sqrt{ \Delta^2 + (n_L + \gamma) (2\hbar eB_{\zeta} v_{\xi} v_{\eta})} + \frac{1}{2} g_{\text{eff},\zeta} \mu_B B_{\zeta},  n \ge 0 ,
\end{split}
\end{equation}
where $\gamma$ accounts for the effect of the Berry phase of the subbands and $B_{\zeta}$ represents the Zeeman shift. 
$B_{\zeta}$ is the magnitude of the magnetic field along $[112]$. $g_{\text{eff},\zeta} = g_0 + 2 g'$ consists of the usual spin Zeeman effect and the anomalous correction from orbital motion. 

Since the g-factor of  topological surface states near the bottom of the lowest conduction subband is larger than that of higher Landau levels, the degeneracy of the first Landau level is lifted due to spin splitting. As a result, both the $\nu=1$ and $\nu=2$ quantum Hall plateaus can be observed in \CdAs~films of all thicknesses studied in our experiments. 

Next, we discuss the emergence of the $\nu=3$ quantum Hall state which indicates an anomalous enhancement of the g-factor with decreasing film thickness. 
To understand the thickness dependence of the g-factor, we consider two quantum confinement contributions to the g factor as follows. First, as the film thickness decreases, the confinement energy splitting between quasi-2D subbands increases. As a result, the magnitude of the correction to the g-factor, which is inversely proportional to the energy gap between subbands, decreases. 
This has been observed in {\CdAs} films with thickness varying from $12$nm to $100$nm in, for example, Ref. \cite{Uchida2017}, where the substrate and the doping level are different from our system.
Secondly, for an ultrathin film of a topological material, hybridized bands with surface-like states near the chemical potential can play an important role in transport. 
Since these surface-like states break parity, they carry symmetry lower than that of the bulk bands. 
Consequently, more subbands are allowed by symmetry to contribute to the perturbation correction of the $g$-factor based on Roth's formula \cite{Roth1959,Winkler2003}. 
Furthermore, for our thin films of {\CdAs}, the energy of the valence subband closest to the hybridized massive Dirac bands, $v_2$ in Fig. \ref{Figure 3} (e), can exhibit a non-monotonic thickness dependence. (This can be inferred from non-monotonic dispersion of the $\Gamma_6^-$ bulk band along $\Gamma$-$Z$ in Refs. \cite{Wang2013} and \cite{Baidya2020}.) 
Therefore, the g-factor can inherit a complicated thickness dependence on the energy gaps between hybridized bands at chemical potential and other nearby subbands. A detailed derivation of the $g$-factor for the ($112$) {\CdAs} thin film is provided in Sec. VII of the Supplementary Material \cite{Xiao_supp}.

As shown in Fig. \ref{Figure 3} (f), in thick films, $g_\text{eff}$ approaches its large value ($\sim 20$) in the bulk. When the films get thinner than 10 nm, $g_\text{eff}$ quickly quenches with a knee to a value of about 5 at the thickness of 5 nm. However, when the thickness is between 10 nm and 15 nm, the dependence of $g_\text{eff}$ on thickness is non-monotonic. This feature is consistent with our experimental observation of a signature of the $\nu=3$ state in the 10 nm thick film. In addition, besides the correction from the nearby subbands, these features (knee, quench, slight rising) are highly dependent on the specific band dispersion which can be affected by details of the hybridization, the confinement potential, and the substrate. These effects, especially the influence of the substrate, likely play an important role in the differences between our data and that in a previous report which used \CdAs~films grown on a different substrate \cite{uchida2017quantum}. 


Finally, we take a detour to gain additional insights into the development of the IQHE in \CdAs~thin films when the chemical potential is near the CNP. We do this by addressing ``vertical'' line cuts of the LL fan diagram shown in Fig. 1 (d), recalling that the data in that figure were generated by sweeping the gate voltage at fixed magnetic field. In moderately disordered 2D electron gases,  one expects to observe the onset of SdH oscillations once the magnetic field is strong enough so that the LL gap overcomes the disorder broadening. At even stronger magnetic fields where the LL separation also exceeds the thermal activation energy, the SdH oscillations transition to the IQHE. However, Fig. 1(d) clearly indicates that this is not the case when the chemical potential is near the CNP. 

\begin{figure*}
\includegraphics[width=0.9\textwidth]{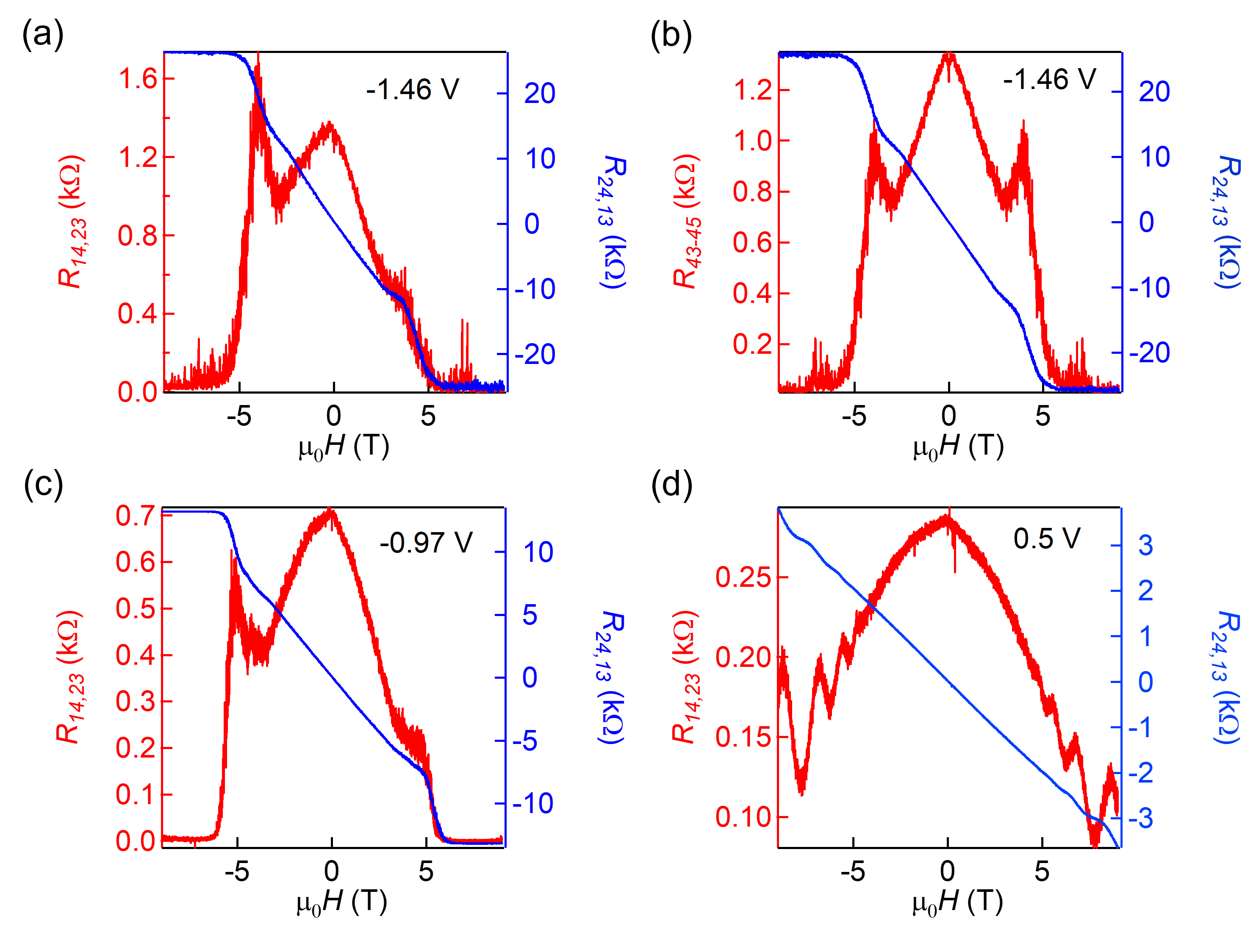}
\caption{\label{Figure 4} Magnetic field dependence of various four-probe resistances at different values of \Vg~in the 15 nm thick film shown in Fig. 1(a). (a) As measured longitudinal resistance ($R_{14,23}$, red, left axis) and transverse resistance ($R_{24,13}$, blue, right axis) at $V_g = -1.46$ V. (b) Field-symmetrized ($R_{14,23}$) and antisymmetrized ($R_{24,13}$) data shown in (a). (c) $R_{14,23}$ (red, left axis) and $R_{24,13}$ (blue, right axis) at $V_g = -0.97$ V. (d) $R_{14,23}$ (red, left axis) and $R_{24,13}$ (blue, right axis) $V_g = 0.5$ V, showing SdH oscillations. 
}
\end{figure*}

To see this behavior in more detail, we now measure the longitudinal and Hall resistance in the 15 nm thick sample while sweeping the magnetic field at fixed gate voltage (in the range $-1.46 ~\mathrm{V} \leq V_g \leq 0.5$~V). For convenience, we only report measurements using one set of four-probes (terminals 1,2,3,4 in the device shown in Fig. 1 (a)) without doing the standard averaging used for a the van der Pauw geometry. The longitudinal resistance is measured as $R_{14,23}$ and the Hall resistance is $R_{24,13}$. Figures \ref{Figure 4} (a) and \ref{Figure 4} (c) show that the magnetic field dependence of the longitudinal resistance does not display any SdH quantum oscillations in this regime of chemical potential. Instead, over most of the magnetic field range, we only observe a background negative magnetoresistance that culminates in a negligible longitudinal resistance upon reaching the $\nu = 1$ quantum Hall state. The magnetic field dependence of the Hall resistance is also quite linear and monotonic until the fully quantized Hall plateau is reached. As the magnetic field sweeps through the value for the $\nu = 2$ state and enters the fully quantized $\nu = 1$ state, we observe a jump in longitudinal resistance which is highly asymmetric in magnetic field direction. This field asymmetry is likely the result of an admixture of the large Hall contribution into the longitudinal resistance. We can remove this cross talk by the standard field symmetrization and antisymmetrization of the longitudinal and Hall resistances, respectively, as shown in Fig. \ref{Figure 4} (b). Although the magnetic field sweep shows a direct transition from an insulator to a quantum Hall liquid state for chemical potential near the CNP, this picture changes dramatically at higher values of the chemical potential: Fig. \ref{Figure 4} (d) shows that when \Vg $>>$ \VgO, we observe normal SdH oscillations at higher LL indices. 

A simple explanation of the anomalous behavior near the CNP is that reduced screening in this regime leads to disorder broadening of the LLs, resulting in suppressed quantum oscillations at low magnetic fields. The degree of LL broadening could be severe enough that the quantum Hall effect is only possible at the highest magnetic field when a single LL is filled. This could be an example of the class of magnetic field-driven insulator to quantum Hall liquid transitions originally reported in early studies of moderately disordered semiconductor-based 2D electron gases at low density \cite{Jiang_PhysRevLett.71.1439,Song_PhysRevLett.78.2200,Smorchkova_PhysRevB.58.R4238}. A more complete examination of this conjecture will require additional studies of the quantum transport as a function of temperature to establish whether the state at zero magnetic field is indeed indicative of a strongly localized insulator. As far as we know, a direct magnetic field induced insulator to quantum Hall liquid transition has not yet been reported in other Dirac materials such as graphene.

In summary, we have mapped out the behavior of SdH oscillations and the IQHE in quantum confined \CdAs~thin films as a function of magnetic field and chemical potential. When the chemical potential is above the CNP, we observe clear quantum oscillations and quantum Hall plateaus at magnetic fields larger than 4 T. Under these conditions, quantum Hall states with filling factors $\nu = 1, 2, 4, 6$ are well-developed in 10 to 15 nm thick films. As the film thickness approaches 10 nm, a signature of the $\nu=3$ quantum Hall state begins to emerge due to an enhancement of the g-factor (and thus larger Zeeman splitting). Tight-binding calculations show that quantum confinement is likely responsible for this enhanced g-factor. The g-factor in ultrathin films may also be larger due to the correction from nearby subbands. These new insights change earlier expectations that the g-factor in \CdAs~would become monotonically smaller as the film thickness decreases. The combination of experiment and theory used in this Letter gives us a comprehensive understanding of the band structure and the thickness dependence of the g-factor in \CdAs~thin films under quantum confinement. We anticipate that our results will motivate more thorough studies of the IQHE in strongly quantum confined topological semimetal thin films. Finally, we also showed that studies of quantum transport in \CdAs~thin films with the chemical potential near the CNP could provide a new Dirac platform for studying the insulator to quantum Hall liquid transition in strongly localized electron systems.

\begin{acknowledgments}
This project was principally supported by the Institute for Quantum Matter under DOE EFRC grant DE-SC0019331. We thank Brad Ramshaw for valuable discussions.  

\end{acknowledgments}

\providecommand{\noopsort}[1]{}\providecommand{\singleletter}[1]{#1}%

\newpage

\end{document}